\documentclass[showpacs,superscriptaddress,preprintnumbers,amsmath,amssymb,onecolumn,nofootinbib,12pt]{revtex4}
\usepackage{graphicx}
\usepackage{dcolumn}
\usepackage{bm}

\usepackage{xcolor}

\usepackage[colorlinks,
            linkcolor=purple,
            anchorcolor=purple,
            citecolor=green
            ]{hyperref}

\begin{document}

\title{Universal Critical Exponents of Non-Equilibrium Phase Transitions from Holography}

\author{Hua-Bi Zeng}\email{hbzeng@yzu.edu.cn}
\affiliation{Center for Gravitation and Cosmology, Yangzhou University, Yangzhou 225009, China}
\affiliation{College of Physics Science and Technology, Yangzhou University, Yangzhou 225009, China}

\author{Hai-Qing Zhang}\email{hqzhang@buaa.edu.cn}
\affiliation{Department of Space Science \& International Research Institute for Multidisciplinary Science,
Beihang University, Beijing 100191, China}

\begin{abstract}
We study the critical exponents in the universal scaling laws of a holographic non-equilibrium steady state nearby its critical point of phase transition, which is driven by an AC electric field sitting in the boundary of the bulk. The applied electric filed drives the initial superconducting state into a non-equilibrium steady state with vanishing condensate as its amplitude is greater than a critical value. In the vicinity of the non-equilibrium critical point, we numerically calculate the six static and one dynamical critical exponents, and find that they have similar values to those in equilibrium systems within numerical errors.

\end{abstract}

\pacs{ 11.25.Tq, 74.25.N,  74.40.Gh}

\maketitle
\pagebreak

\section{Introduction}
\label{sec:intro}
 One of the most intriguing features in equilibrium continuous phase transition is the universal scaling behavior near the critical point, which groups various critical phenomena into universality classes, i.e., systems lie in one universality class share the same scaling behavior \cite{HH:1977}. From the point of view of modern physics, the concept of universality has its origin in the renormalization group \cite{Wilson}, that the universality roots in the long-range correlations in the system no matter the microscopic detail is.  However, little is understood about the general aspects of {\it non}-equilibrium systems. A full classification of the universality classes in non-equilibrium phase transitions is still lacking because of the violation of the detailed balance \cite{Hinrichsen:2005}, thus the fluctuation-dissipation theorem cannot be applied \cite{HH:1977,Racz:2004}. However, it is believed that even for a system far away from equilibrium, the concepts of scaling and universality can still be applied to the non-equilibrium phase transition. There are many attempts in condensed matter physics trying to explain the power law correlations presented in non-equilibrium dynamics \cite{Odor:2004}. These studies reveal a very close connection between equilibrium and non-equilibrium critical phenomena. Thus, investigating the similarities or differences between equilibrium and non-equilibrium systems may help us to understand the essence of non-equilibrium dynamics.
 
Non-equilibrium steady state (NESS) is different from the equilibrium state that NESS has presence of fluxes in the systems, either by the boundary condition or by bulk driving fields. A well-known example of NESS is the Rayleigh-B\'enard experiment that the horizontal layers of viscous fluid sitting between two heat baths with temperatures $T_1$ and $T_2$. For $T_1 = T_2$ this system relaxes into a quiescent equilibrium state while a small difference of the temperature, e.g., a small $\delta T=|T_1-T_2|$ will have a NESS since energy flux is flowing through the system. We are going to focus on NESS regime in non-equilibrium state. Increasing $\delta T$ will make the system have more sophisticated structures which will not be explored in this paper, interested readers may refer to \cite{cross:1993}. 

At equilibrium states, the second-order phase transitions result from the long-range correlations, regardless of the original short-range interactions. Interestingly, there is usually a collective behavior over large scales in a strongly coupled complex system \cite{Henkel}. Therefore, it is of particular interest to study the long-range behaviors of strongly coupled non-equilibrium dynamics, and to see whether it will have similar behavior to the equilibrium long-range correlations. Except for the numerical Monte-Carlo simulation \cite{Kurt} and lattice gauge theories \cite{Rothe}, strongly coupled field theory is notorious for the difficulty to be solved analytically since perturbative methods are impossible to be implemented in the strongly coupled regime. Fortunately, in recent years people from high energy physics have invented a holographic approach to study the strongly coupled field theory from the weakly coupled gravity, which is dubbed AdS/CFT correspondence \cite{Maldacena:1997re}. 
 
In the limit of large gauge group rank $N_c$ and large 't Hooft coupling $\lambda$, the AdS/CFT correspondence can study the strongly coupled field theory from the weakly coupled gravity\cite{Maldacena:1997re,Gubser:1998bc,Witten:1998qj,Aharony:1999ti}. Its application in equilibrium/non-equilibrium dynamics has been investigated in various ways. For instance, AdS/CFT correspondence has been adopted in the study of hot QCD and strongly coupled quark-gluon plasma \cite{CasalderreySolana:2011us}; the non-equilibrium dynamics of superconducting order parameter after quench \cite{Murata:2010dx,Bhaseen:2012gg}; topological defects formation in Kibble-Zurek mechanism \cite{Chesler:2014gya,Sonner:2014tca}; time evolution of non-local entanglement observables \cite{Bai:2014tla,Erdmenger:2017gdk}; energy flows between two heat baths \cite{Bhaseen:2013ypa}, and etc. Interested readers may refer to the review papers \cite{Hubeny:2010ry,Chesler:2013lia}. 

In this paper, we are going to investigate the scaling laws nearby the critical point of holographic non-equilibrium steady states, which are driven by a sinusoidal applied AC electric field. The universal scaling laws nearby the critical point in non-equilibrium dynamics were already partially studied in \cite{Chesler:2014gya,Sonner:2014tca,Nakamura:2012ae,Matsumoto:2018ukk}.  At the initial time, the state is in a static superconducting phase which can be obtained from the holographic construction of the charged scalar model \cite{Gubser:2008,HHH:2008}. After the initial time, we add a sinusoidal AC electric field on the boundary of the spacetime to drive the system away from the static phase \cite{Li:2013fhw}. The superconducting order parameter will decrease dramatically according to the amplitude $E$ of the applied electric field, while we have fixed the frequency of the electric field. Therefore, we can regard the amplitude as the controlling parameter which drives the system away from the initial static state.  Eventually the final state will saturate into a superconducting non-equilibrium steady state or normal steady state, depending on the strength of the amplitude. It is found that  there exists a critical value of the amplitude $E_c$, beyond which the system will finally become a normal state. Based on this non-equilibrium phase transition, we investigate the scaling laws nearby the critical point $E_c$. We numerically explore the six static critical exponents, i.e. $(\alpha, \beta, \gamma, \delta, \eta, \nu)$ and one dynamical critical exponents $z$. The details of the computation can be found in Section \ref{sec:CE}. We found that the critical exponents are numerically consistent with those in mean field theory within numerical errors, which indicates that the holographic system on the boundary in large N limit does not have great discrepancy from the mean field theory. These results match those in the previous holographic studies. 

The paper is arranged as follows: The holographic background is constructed in Section \ref{sec:bg}, while the numerical computations of the critical exponents are given in Section \ref{sec:CE}; We draw the conclusions and discussion in Section \ref{sec:cd}. In the Appendix \ref{sec:app} we briefly review the critical exponents we considered in mean field theory.

\section{Holographic non-equilibrium phase transition}
\label{sec:bg}

The action we adopt is a $U(1)$ gauge field $A_\mu$ coupled with a complex scalar field $\Psi$ (for simplicity, we work in the probe limit by ignoring the backreaction of the scalar fields and gauge fields to the gravity), 
\begin{equation}
S = \int d^{4}x \sqrt{-g} \left( -\frac{1}{4} F_{\mu\nu} F^{\mu\nu} - |\partial_\mu \Psi - i A_\mu \Psi|^2 - m^2 |\Psi|^2 \right).
\end{equation}
where $F_{\mu\nu}=\partial_\mu A_\nu-\partial_\nu A_\mu$ is the gauge field strength while $m^2$ is the mass square of the scalar field.  The gravity background is the neutral $\textrm{AdS}_4$ planar black hole, with the metric in Eddington-Finkelstein coordinates as
\begin{equation}\label{metric}
ds^2 = \frac{1}{z^2} \left( -f(z) dt^2 - 2 dt dz + dx^2 + dy^2 \right),
\end{equation}
with $f(z) = 1 - z^3$ (we have scaled the AdS radius as $L=1$). Therefore,  $z=1$ is the location of horizon while $z = 0$ is the boundary where the field theory lives. Dynamics of the system is governed by the following time dependent equations of motion:
\begin{eqnarray}
\label{eompsi}
\partial_t \partial_z \Phi - i A_t \partial_z \Phi - \frac12 \Bigl[  f \partial_z^2 \Phi + f' \partial_z \Phi
+ i \partial_z A_t \Phi - z \Phi - A_x^2 \Phi \Bigr] &=& 0,
\\
\label{eom2}
\partial_t \partial_z A_t + 2 A_t |\Phi|^2 - i f (\Phi^* \partial_z \Phi - \Phi \partial_z \Phi^*) 
+ i (\Phi^* \partial_t \Phi - \Phi \partial_t \Phi^*) &=& 0,
\\\label{eomax}
\partial_t \partial_z A_x - \frac12 \left[ \partial_z ( f \partial_z A_x) - 2 A_x |\Phi|^2 \right] &=& 0,\\
\label{eom1}
\partial_z (\partial_z A_t) - i (\Phi^* \partial_z \Phi - \Phi \partial_z \Phi^*) &= &0.
\end{eqnarray}
with the ansatz that: $\Phi= \Psi(t, z)/z, \, A_t=A_t(t, z), \, A_x=A_x(t, z)$ and $A_z=A_y=0$. The above four equations in fact satisfy the following constraint equation
\begin{eqnarray}\label{conservation}
\frac{d}{dt}\text{Eq.\eqref{eom1}}-\frac{d}{dz}\text{Eq.\eqref{eom2}}\equiv-2i\left(\text{Eq.\eqref{eompsi}}\times\Phi^*-c.c.\right)
\end{eqnarray}
where $c.c.$ represents complex conjugation. The constraint equation Eq.\eqref{conservation} actually originates from $\nabla_\mu\nabla_\nu F^{\mu\nu}\equiv0$, implying the conservation of current.
The asymptotic expansions of the fields near the boundary are (we have set the mass square $m^2 = -2$ without loss of generality),
\begin{equation}
 \Psi(t,z)|_{z\to0} = \Psi^{(1)}(t) z + \Psi^{(2)}(t) z^2, \quad A_\mu(t,z)|_{z\to0} = a_\mu(t) + b_\mu(t) z.
\end{equation}
According to the holographic dictionary, $\Psi^{(1)}$ is regarded as the source term of the boundary scalar operator $O$ while $\Psi^{(2)}$ as the vacuum expectation $\langle O\rangle$; The coefficients $a_\mu$ and $b_\mu$ are corresponding to the velocity $v_\mu$ and current $J_\mu$ of the boundary field, respectively. It is worth mentioning that the above four coefficients all depend on time direction as we study the non-equilibrium dynamics of the system.

\begin{figure}[t]
\begin{center}
\includegraphics[trim=0.7cm 6.cm 1.4cm 7.2cm, clip=true, scale=0.4]{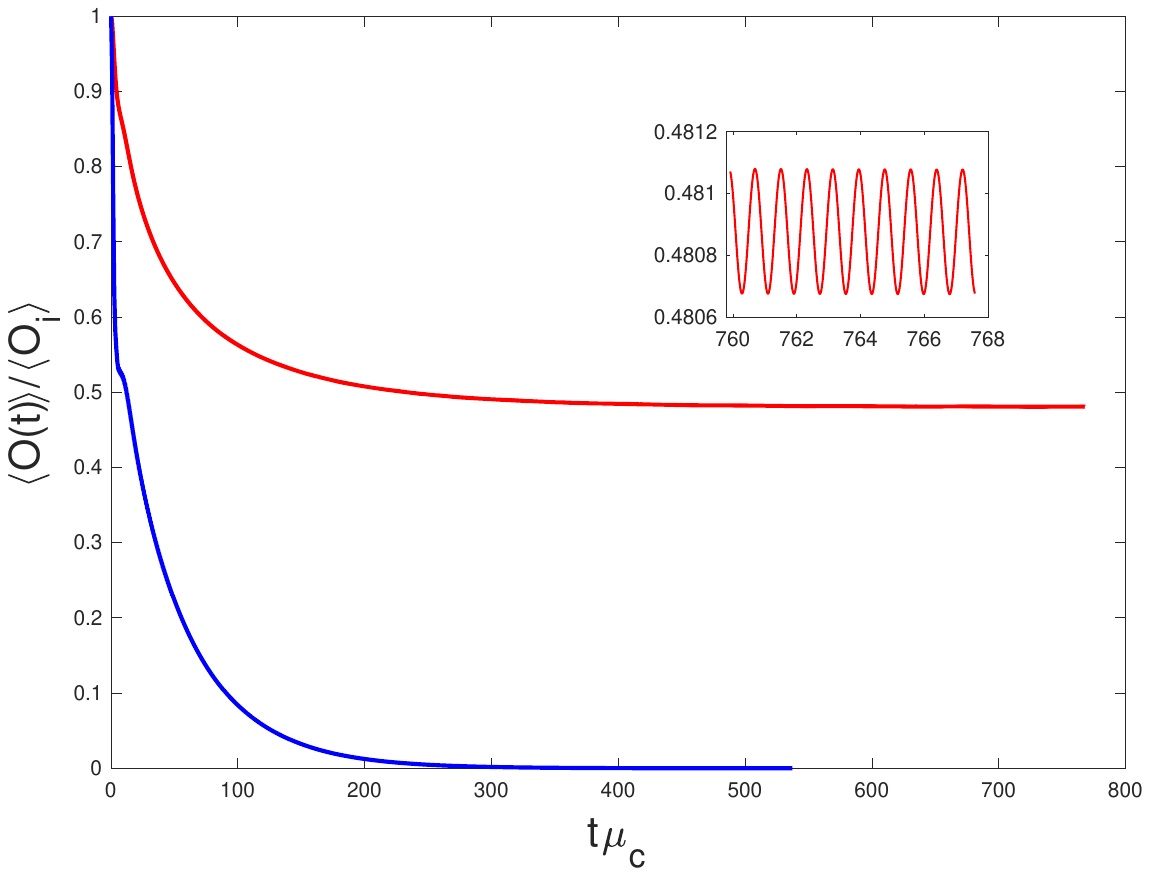}
\includegraphics[trim=0.8cm 6.5cm 2.cm 7.6cm, clip=true, scale=0.43]{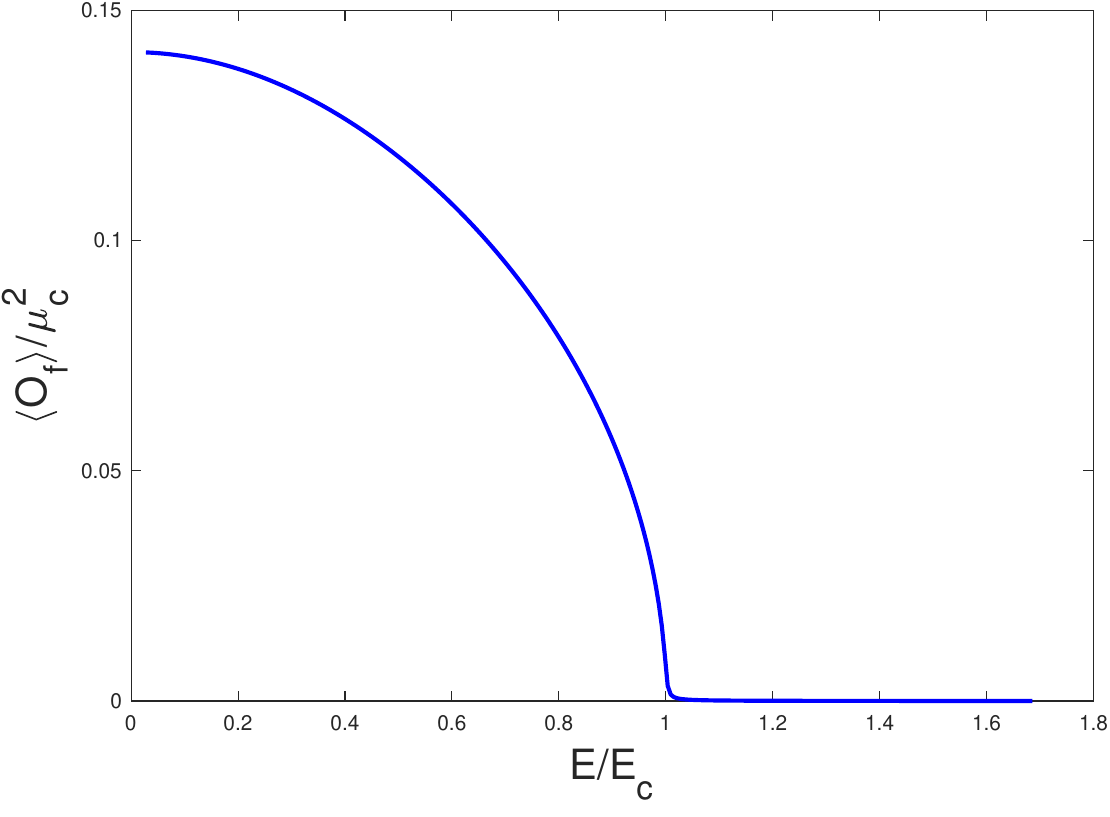}
\caption{(Left panel) Time evolution of the ratio between the condensate $\langle O(t)\rangle$ and the initial condensate $\langle O_i\rangle$. The red and blue lines are corresponding to the amplitudes $E=0.8316E_c$ and $E=2.4948E_c$ respectively. The inset plot shows the tiny oscillations of the order parameter for $E=0.8316E_c$ in the late time, which indicates the non-equilibrium steady state; (Right panel) The relation between the final average condensate $\langle O_f\rangle$ in the late time and the amplitude $E$.  As $E<E_c$ the final state is in the ordered phase with finite condensate $\langle O_f\rangle$, while $E>E_c$ the state is in the disordered phase with vanishing condensate.  The critical value for the phase transition is $E_c\approx 2.1778\mu_c^2$.  } \label{OT}
\end{center}
\end{figure}

One may notice that in the Eq.\eqref{eomax} the gauge field component $A_x$ can vanish independently, however, in order to investigate the non-equilibrium dynamics of the system in response to the external driving force, we turn on $A_x$ and impose its $z = 0$ boundary condition as \cite{Li:2013fhw}:
\begin{equation} \label{Ax}
A_x(t, z=0) = \frac{E \sin(	\Omega t)}{\Omega}.
\end{equation}
 Thus on the boundary the electric field along $x$-direction is $E_x(t) = \partial_t A_x = E \cos(\Omega t)$, in which $E$ and $\Omega$ are the amplitude and frequency of the applied electric field respectively.\footnote{The non-linear transport coefficients of this model has been intensively studied in \cite{Zeng:2016api,Zeng:2016gqj}. } In the static case (time-independent), the system has two kinds of phases: One is the disordered phase in high temperature regime without any condensates of the order parameter; The other one is the ordered phase with scalar condensates as the order parameter in low temperature regime. The temperature of the black hole is $T=3/(4\pi z_h)$, in which $z_h$ is the horizon and we have scaled it to be $z_h\equiv1$.  From static holographic superconductors~\cite{HHH:2008}, the critical point for the phase transition is $\mu_c \approx 4.07$,  thus the critical temperature is $T_c \approx 0.06\mu_c$.  We assume that the initial condition at $t = 0$ is the static solution with a fixed chemical potential $\mu=1.1056\mu_c$, i.e., the system is in the ordered phase/superconducting phase. After the initial time, we then turn on the applied electric field Eq.\eqref{Ax} to drive the system away from the equilibrium state. The system will finally saturate into a non-equilibrium steady state after certain time, depending on the amplitude $E$ and the frequency $\Omega$ of the applied electric field. We need to stress that after a long enough time, the ultimate state is {\it not} an equilibrium state, but rather a non-equilibrium steady state which has very tiny steady oscillations in the order parameter because of the sinusoidal applied electric field. From the inset plot of left panel in Fig.\ref{OT}, we see that at late time there are tiny steady oscillations of the order parameter compared to its average value, in the order of $\approx 10^{-4}$. Although the final order parameter is not exactly an constant, within the numerical errors we can still use the average value of it to compute the properties of the system \cite{Li:2013fhw}. For instance, we can approximately make use of $\langle O(t)\rangle/\langle O_i\rangle\approx0.4809$, where $\langle O_i\rangle$ is the initial value of the condensate, as the value of the order parameter in the late time  for the red line.  In the numerics, we fix the frequency of the applied electric field to be $\Omega=3.8594\mu_c$ while varying the amplitude $E$ to drive the system into various non-equilibrium states and ultimately various non-equilibrium steady states.  The EOMs are solved by the fourth order Runge-Kutta method in $t$-direction and Chebyshev spectral methods in $z$-direction.

On the left panel of Fig.\ref{OT} we show the time evolution of the condensate for two different amplitudes $E=0.8316E_c$ (the red line) and $E=2.4948E_c$ (the blue line). The vertical axis is the ratio between the condensate $\langle O(t)\rangle$ and the initial static condensate $\langle O_i\rangle$. We see that the condensates decrease quickly as we turn on the applied electric field after $t=0$. The condensate with larger $E$ decreases more rapidly than the one with smaller $E$; In the late time for the non-equilibrium steady state,  the condensate will become flat with very tiny oscillations as we have explained above. Therefore, from now on we will use the average value of the condensate to represent the late time condensate  $\langle O_f\rangle$ of the order parameter. The final condensate  $\langle O_f\rangle$ is larger if $E$ is smaller as illustrated in Fig.\ref{OT} by the red line and blue line with $(E=0.8316E_c, \langle O_f\rangle\approx 0.4809\langle O_i\rangle)$ and $(E=2.4948E_c, \langle O_f\rangle \approx 0)$ respectively. The right panel of Fig.\ref{OT} shows the final condensate $\langle O_f\rangle$ with respect to the amplitude $E$, and we find that the critical point for the phase transition from finite $\langle O_f\rangle$ to vanishing $\langle O_f\rangle$ is around $E_c\approx 2.1778\mu_c^2$. Therefore, we see that the amplitude parameter $E$ of the applied electric field can literally drive the original superconducting/ordered phase into a normal/disordered phase, which is a non-equilibrium steady state. 

\section{Critical Exponents in Non-Equilibrium Dynamics}
\label{sec:CE}

In the equilibrium theory, one can use the critical exponents of the power law to study the critical behavior of the phase transition near the critical point. Usually there are six {\it static} critical exponents $(\alpha, \beta, \gamma, \delta, \eta, \nu)$ and one {\it dynamical} critical exponents $z$ respectively\footnote{One should not confuse the dynamical exponents $z$ with the radial direction $z$ in this paper. Indeed, they can be easily distinguished from the contexts.}. In the Appendix \ref{sec:app}, we briefly review the critical exponents in equilibrium dynamics. It will be interesting to see whether these critical exponents or scaling laws still hold in the non-equilibrium dynamics, or to see how much they will deviate from the equilibrium case.  In the following we will study these critical exponents near the critical point of non-equilibrium phase transition from holography, which corresponds to strongly coupled systems on the boundary.  As we mentioned above, the amplitude $E$ in Eq.\eqref{Ax} can be regarded as the external source which drives the system away from the critical point $E_c$, therefore, the non-equilibrium phase transition will have certain scaling laws with respect to the difference $\epsilon_E=1-E/E_c$ as we have expected. We numerically calculate these critical exponents in the following subsections. \footnote{The paper \cite{Matsumoto:2018ukk} also computed the critical exponents of a non-equilibrium phase transition from holography. However, the setup there was different from ours that they worked in a D3-D7 brane system. Moreover, in their paper the authors did not study the critical exponent $\eta$ and the dynamical critical exponent $z$. }

\subsection{Static Critical Exponents}

$\bullet~\alpha=0$: From the Appendix \ref{sec:app}, the critical exponent $\alpha$ is related to the heat capacity of the system as $C\propto|\epsilon_E|^{-\alpha}=|1-E/E_c|^{-\alpha}$. Following the arguments in \cite{Maeda:2009wv}, we can see that in the disordered phase or the phase with $E>E_c$, there is no condensate of the scalar fields. Therefore, the heat capacity of the system is the heat capacity of the black hole. Thus, as we approach the critical point of the non-equilibrium phase transition, the heat capacity converges to a constant since we worked in the probe limit. Therefore, we can deduce that $\alpha=0$. 
 
\begin{figure}[h]
\begin{center}
\includegraphics[trim=0.5cm 6.7cm 1cm 7.7cm, clip=true, scale=0.4]{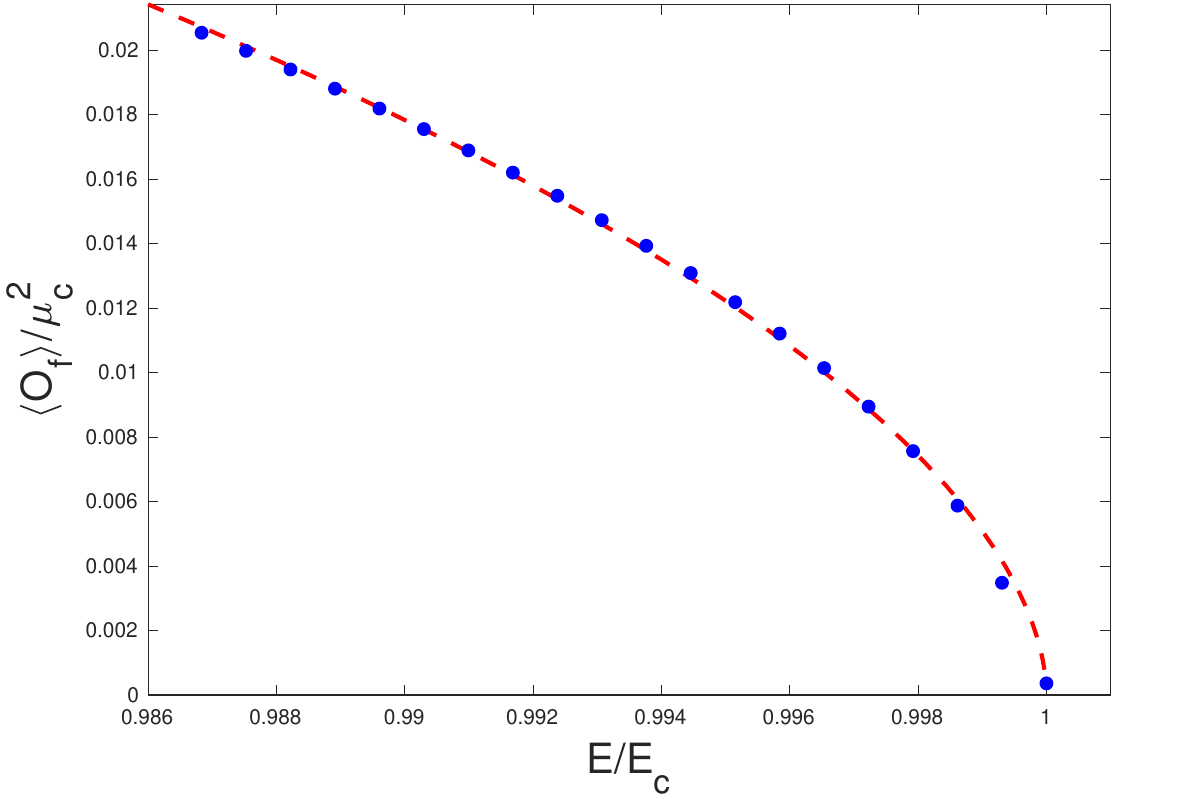}
\caption{Relation between $\langle O_f\rangle$ and $E$ in the vicinity of $E_c$. The  blue dots are the numerical values while the red dashed line is the fitted curve with  $\langle O_f\rangle/\mu_c^2\approx 0.2189\times|1-E/E_c|^{1/2}$.} \label{EO}
\end{center}
\end{figure}

 $\bullet~\beta=1/2$: The critical exponent $\beta$ can be read off from the relation between the order parameter $\langle O_f\rangle$ with respect to $\epsilon_E$ nearby the critical point such that $\langle O_f\rangle/\mu_c^2\propto|1-E/E_c|^{\beta}$. By increasing $E$  in the Eq.\eqref{Ax}, the order parameter in the final non-equilibrium steady state decreases and vanishes at the critical point $E_c\approx 2.1778\mu_c^2$. Fig.\ref{EO} shows the relation between $\langle O_f\rangle$ and $E$ nearby the critical point, the  blue dots are the numerical results while the  red dashed line is the fitted curve, which is roughly  $\langle O_f\rangle/\mu_c^2\approx 0.2189\times|1-E/E_c|^{1/2}$. The numerical dots and the fitted curve match very well. Therefore, we see that $\beta\sim1/2$, which is the same as the value of $\beta$ in the mean field theory.

$\bullet~\nu=1/2$:
To compute the critical exponents $\nu$ we need to solve the perturbative
equations of the system, and then we can get the relation between the correlation length $\xi$ and $\epsilon_E$, such as $\xi\propto|\epsilon_E|^{-\nu}$. Basically $\nu>0$, which indicates the typical divergence of the correlation length near the critical point. The correlation length $\xi$ can be read off from the correlation function (response function) of the order parameter. In the Fourier space the correlation function reads, 
\begin{equation}\label{chiwk}
\chi(\omega, k)=\langle O(\omega, k)O^\dagger(-\omega, -k)\rangle \sim \frac{1}{ic\omega+k^2+1/\xi^2}.
\end{equation}
 where $c$ is a parameter, $\omega$ and $k$ are respectively the frequency and momentum of the transformed Fourier modes. The poles of the correlation function correspond to the system's quasi-normal modes (QNMs) \cite{Kovtun:2005ev}, which can depict the relaxation behavior of the system in the late time.  As we already mentioned, the system will go to steady state in the late time. Therefore, it is more convenient to approximately treat the background fields as time-independent in the late time. In order to study the QNMs of the system, we need to perturb the fields linearly.  The first order perturbations of the fields can be written as $\delta A_t(t,z,x)\to  e^{-i\omega t+ikx}\delta A_t(z)$ and $\delta \Phi(t,z,x)\to e^{-i\omega t+ikx}\delta\Phi(z)$.\footnote{We did not perturb the field $A_x$ when we calculated the QNMs, since physically $A_x$ plays the role of the external driving force in the system. On the other hand, QNMs are the first order perturbations which respond to the system. Therefore, it is more physical to treat $A_x$ as the background field which does not contribute fluctuating modes to QNMs. } The equations of motions for the first order fluctuations of the fields read, 
\begin{eqnarray}\label{dAt}
\omega  {\delta{A_t}}'+i  \left(k^2+2 \Phi^2\right)\delta{A_t}+4 i {A_t} \Phi  \delta\Phi &=&0, ~~~~~~\\
\left(iA_t'-z-(A_x-k)^2\right)\delta\Phi +i \Phi {\delta{A_t}}'+2 i \Phi'\delta{A_t}
   +\left(2 i (A_t+\omega) -3z^2\right) {\delta\Phi} '+\left(1-z^3\right) {\delta\Phi} ''&=&0.\label{dPhi}
\end{eqnarray}
where $'$ is the derivative with respect to radial $z$-coordinate. We will calculate the QNMs of the system nearby the critical point, i.e., $E\sim E_c$. Besides, we make use of the average values of the background fields as we calculate the QNMs since the system is in the steady state in the late time as we already mentioned above.  

\begin{figure}[h]
\begin{center}
\includegraphics[trim=0cm 0cm 0cm 0cm, clip=true, scale=0.63]{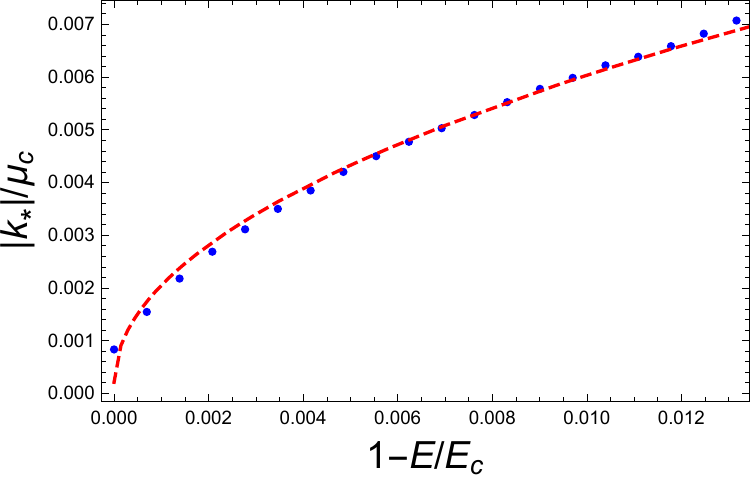}
\includegraphics[trim=1.cm 6.4cm 2.2cm 7.6cm, clip=true, scale=0.37]{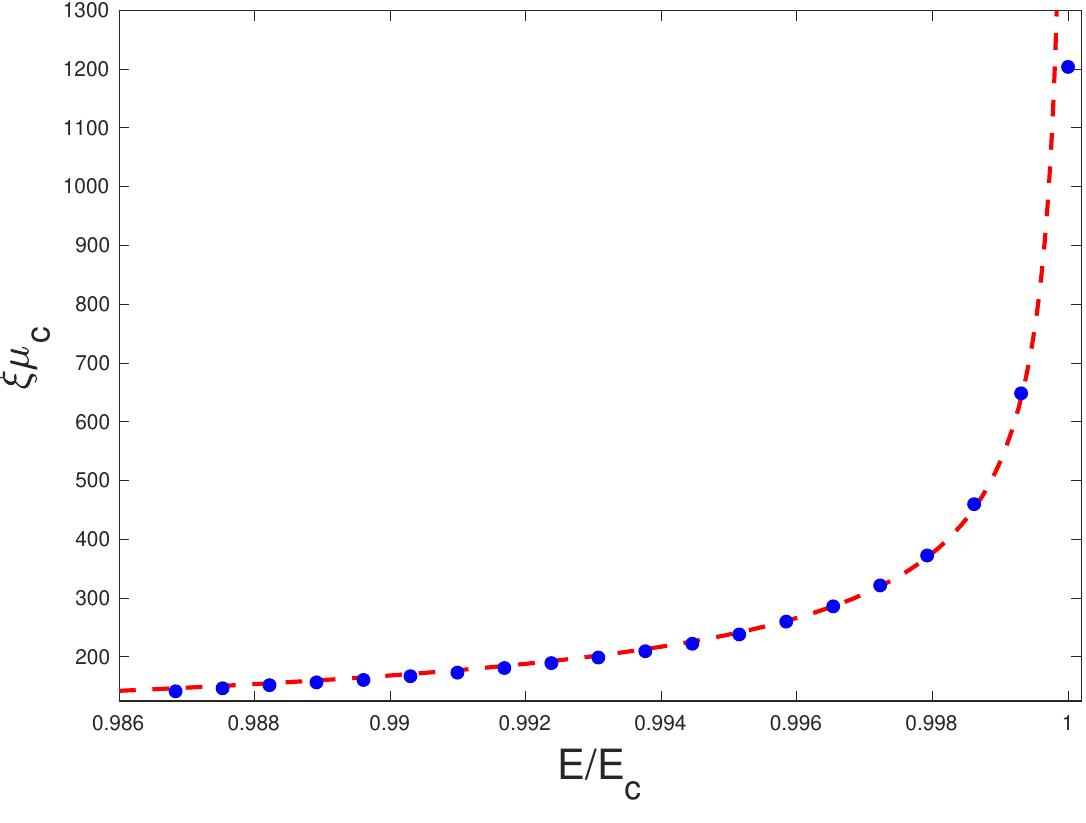}
\caption{(Left Panel) Relation between the lowest modes of momentum $|k_*|$ and $(1-E/E_c)$ with  $|k_*|/\mu_c\approx0.05827\times|1-E/E_c|^{1/2}$; (Right Panel) Relation between the correlation length $\xi$ and $E$ with  $\xi\mu_c\approx17.1615\times|1-E/E_c|^{-1/2}$. For both plots, the dots are the numerical results while the dashed lines are the fitted curves.  } \label{xi}
\end{center}
\end{figure}

From the poles of the response function in Eq.\eqref{chiwk}, we see that the relation between the correlation length $\xi$ and the momentum $k$ can be obtained by solving Eqs.\eqref{dAt} and \eqref{dPhi} and setting $\omega=0$. Therefore, the poles of the response function are located in $k_*^2=-1/\xi^2$. By varying the external parameter $E$, we then select the lowest modes of $k$ which have the negative imaginary parts mostly closing to the real axis, for instance the mode with $E=0.9868E_c$, the lowest $k=-0.007068i \mu_c$.   From the left panel of Fig.\ref{xi} we find that the fitted curve for the numerical data is roughly  $|k_*|/\mu_c\approx0.05827\times|1-E/E_c|^{1/2}$ within numerical errors. Therefore, the correlation length $\xi=1/|k_*|\propto |\epsilon_E|^{-1/2}$, which is shown on the right panel of Fig.\ref{xi}. Thus, we see that near the critical point of the non-equilibrium phase transition $\nu=1/2$, which is the same as the one in the equilibrium field theory.  

\begin{figure}[h]
\begin{center}
\includegraphics[trim=.5cm 6.7cm 1.5cm 7.7cm, clip=true, scale=0.4, angle =0]{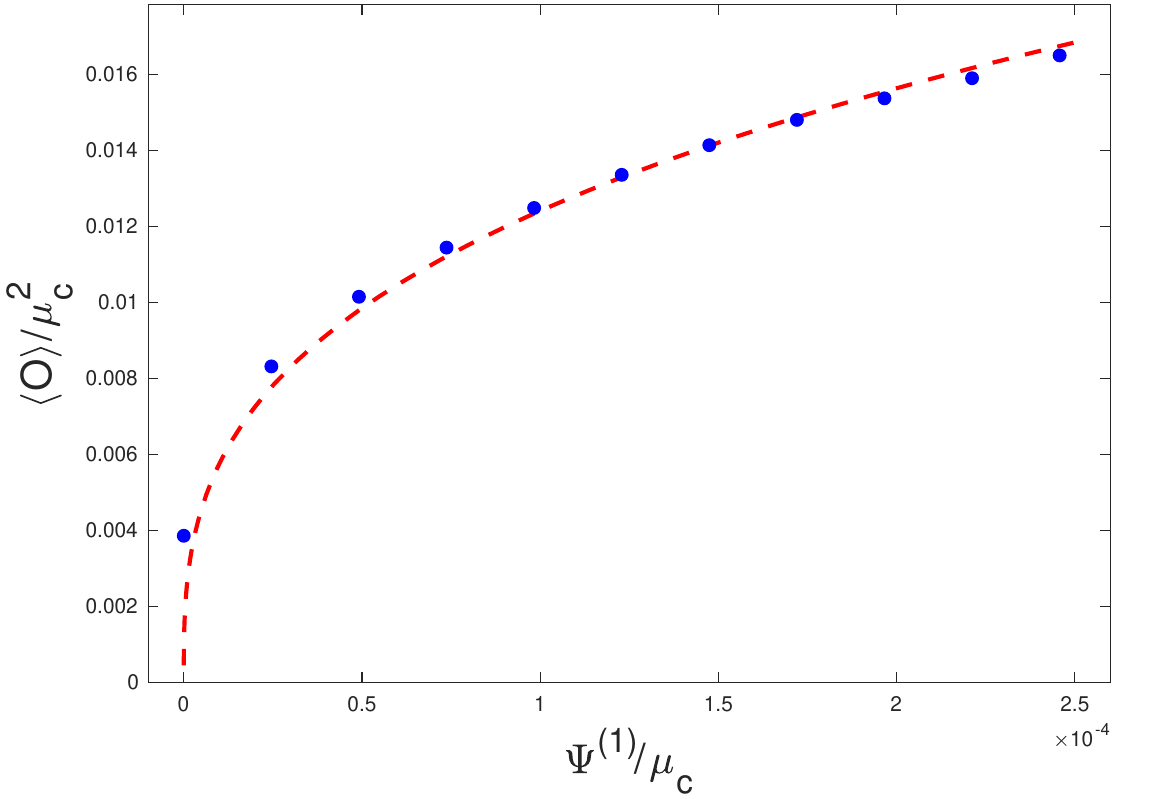}
\caption{Relation between the order parameter and the source, from which we can deduce $\delta= 3$. Dots are the numerical results while the dashed line is the fitted curve, which has relation  $\langle O\rangle/\mu_c^2\approx0.2674\times\left(\Psi^{(1)}/\mu_c\right)^{1/3}$} \label{delta}
\end{center}
\end{figure}

$\bullet~\delta=3$:
The critical exponent $\delta$ can be obtained from the relation between the order parameter and its source term near the critical point, i.e., $\langle O\rangle|_{E\sim E_c}=\Psi^{(2)}\propto \left(\Psi^{(1)}\right)^{1/\delta}$ \cite{Maeda:2009wv}. As we calculate $\delta$, we set $E$ very close to $E_c$ and slightly vary the source $\Psi^{(1)}$ away from zero since we still roughly need the condensate of the order parameter from the spontaneous symmetry breaking. So we keep the source term $\Psi^{(1)}$ in the order of $10^{-4}\mu_c$ which can be seen in Fig.\ref{delta}. In Fig.\ref{delta} we plot the relation between the condensate value of the order parameter and the source, the dots are the numerical results while the dashed line is the fitted curve. Within numerical errors, we find that $\langle O\rangle/\mu_c^2~\big|_{E_c}\approx0.2674\times\left(\Psi^{(1)}/\mu_c\right)^{1/3}$, thus $\delta=3$ as one expected in the equilibrium dynamics.

\begin{figure}[h]
\begin{center}
\includegraphics[trim=1.cm 6.7cm 2.cm 7.6cm, clip=true, scale=0.4]{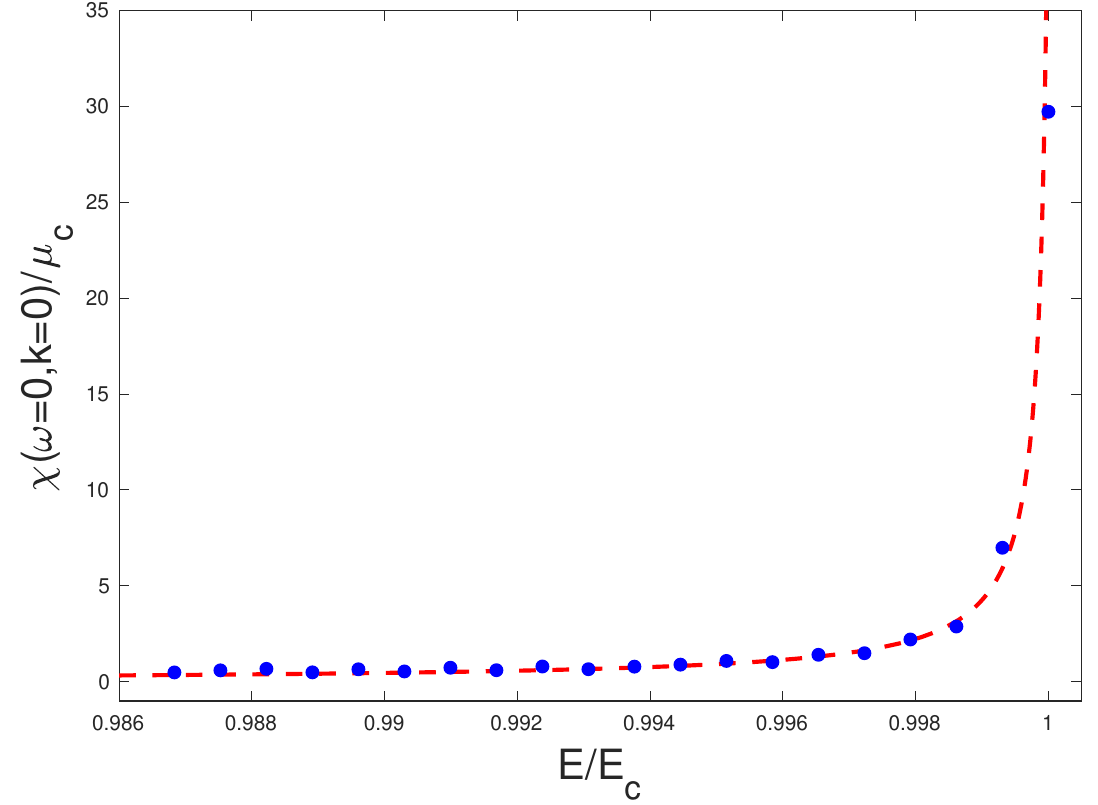}
\caption{ Relation between the response function $\chi(\omega=0, k=0)$ and $E$, from which we can deduce $\gamma=1$. The dots are the numerical results while the dashed line is the fitted curve  with the relation $\chi(\omega=0, k=0)/\mu_c\approx 0.004692\times|1-E/E_c|^{-1}$. } \label{GF}
\end{center}
\end{figure}

$\bullet~\gamma=1$:
The critical exponent $\gamma$ can be read off from $\chi(\omega=0, k=0)\propto|1-E/E_c|^{-\gamma}$. As we know, in the AdS/CFT correspondence the response function is obtained from $\chi(\omega=0, k=0)=\delta\psi_2/\delta\psi_1$, where $\delta \psi_2$ and $\delta\psi_1$ are the perturbations of the order parameter and the source respectively. In the numerical computation we vary the amplitude of the electric field $E$ while fixing $\delta\psi_1|_{z=1}=0.2457\mu_c$, and then to study the ratio $\delta\psi_2/\delta\psi_1$ on the $z=0$ boundary. From Fig.\ref{GF} we can fit  $\chi(\omega=0, k=0)/\mu_c\approx 0.004692\times|1-E/E_c|^{-1}$, thus within numerical errors $\gamma=1$ is the same as that in the equilibrium  field theory. 

\begin{figure}[h]
\begin{center}
\includegraphics[trim=0cm 0cm 0cm 0cm, clip=true, scale=0.65]{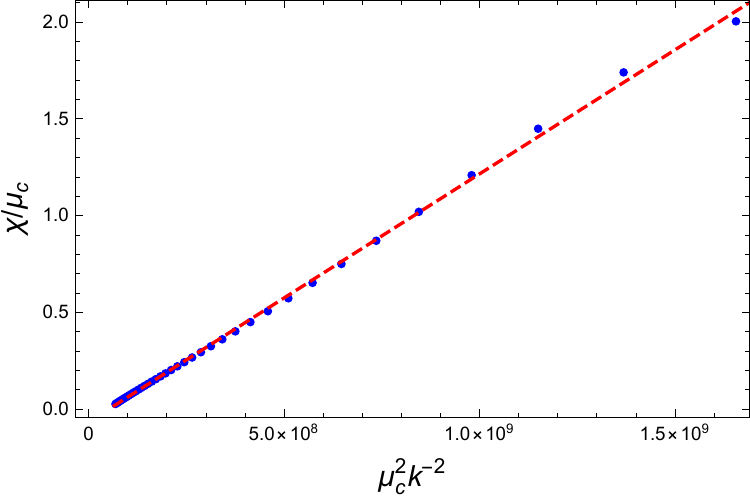}
\caption{ Relation between the response function $\chi$ and $k^{-2}$, from which one can deduce that $\eta=0$. The dots are the numerical results while the dashed line is the fitted curve with the relation $\chi/\mu_c\approx 1.2801\times10^{-9}\mu_c^2 k^{-2}$. } \label{chik}
\end{center}
\end{figure}

$\bullet~\eta=0$:
From $\chi(\omega=0, k)\propto k^{\eta-2}$ one can read off the value of the critical exponent $\eta$. Therefore, in the vicinity of the critical point $E_c$, we slightly change the momentum $k$ (in the order of $10^{-4}\mu_c$) away from zero to calculate the response function $\chi$. The results are shown in the Fig.\ref{chik}. The dots are the numerical results while the dashed line is the fitted curve. Therefore, from Fig.\ref{chik} we see that $\chi$ is linear proportional to $k^{-2}$ with $\chi/\mu_c\approx 1.2801\times10^{-9}\mu_c^2 k^{-2}$, which implies $\eta=0$. This result of $\eta$ is similar to that in equilibrium dynamics. 

Therefore, from the above numerical results of the six static critical exponents we find that they have similar values compared to those in equilibrium dynamics, i.e., $(\alpha, \beta, \gamma, \delta, \nu, \eta)=(0,\frac12,1,3,\frac12,0)$. Moreover, they also satisfy the four identities  \eqref{rush}, \eqref{wido}, \eqref{fish} and \eqref{jose} as shown in the Appendix \ref{sec:app}.

\subsection{Dynamical Critical Exponent $z=2$}

\begin{figure}[h]
\begin{center}
\includegraphics[trim=0.8cm 6.7cm 1.9cm 7.7cm, clip=true, scale=0.36]{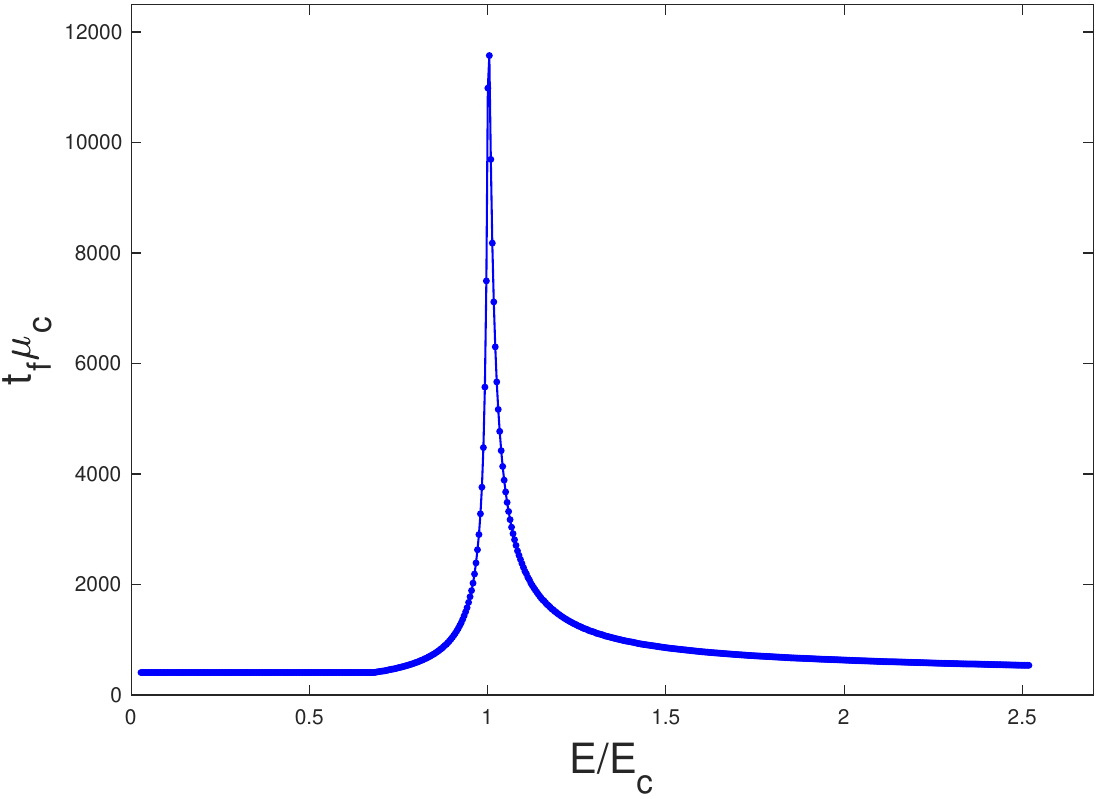}
\includegraphics[trim=0cm 0cm 0cm 0cm, clip=true, scale=0.64]{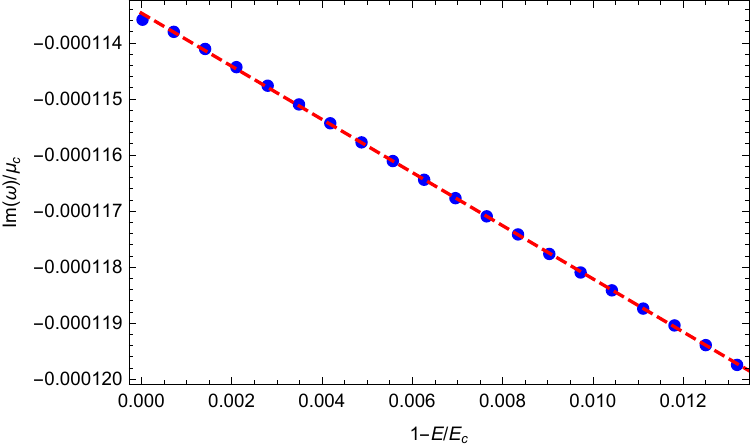}
\caption{(Left Panel) Relation between the saturation time $t_f$ and the amplitude $E$. $t_f$ diverges nearby the critical point $E_c$, which indicates the critical slowing down near critical point; (Right Panel) Linear relation between the imaginary part of the QNMs and $|1-E/E_c|$, from which we can deduce the relation between relaxation time with respect to $|1-E/E_c|$ and hence $z=2$. The dots are the numerical results while the dashed line is the fitted curve with $\text{Im}(\omega)/\mu_c\approx -4.7409\times10^{-4} |1-E/E_c|$ . } \label{relaxationtime}
\end{center}
\end{figure}

After certain time $t_f$, the system will saturate into the non-equilibrium steady states as we discussed above.\footnote{Let's call $t_f$ {\it saturation time} loosely. It will approximately equal to the {\it relaxation time} which will be defined from the QNMs exactly in the following. } In order to find the approximate value of the saturation time $t_f$, we set a threshold that if $\langle O(t+\Delta t)\rangle/\langle O(t)\rangle \lesssim10^{-5}$, where $\Delta t$ is the periodicity of the order parameter in the steady state, we can say that the time $t$ is approximately the saturation time $t_f$ (Please refer to the left panel of Fig.\ref{OT} that $\Delta t\mu_c\approx0.8$ for the red curve).  We show the relation between $t_f$ and the external amplitude $E$ in the left panel of Fig.\ref{relaxationtime}. Nearby the critical point $E_c$, we can see the divergence of the saturation time $t_f$, which indicates the critical slowing down near the phase transition point. 

Dynamical critical exponent $z$ can be read off from the relation $\tau(k=0)\propto\xi^z$ where $\tau$ is the relaxation time. Since we already knew $\xi\propto|1-E/E_c|^{-1/2}$, we need to check the relation between $\tau$ and $|1-E/E_c|$ by varying $E$, hence $\tau(k=0)\propto|1-E/E_c|^{-z/2}$. From the fact that the relaxation time is related to the inverse of the imaginary part of the QNMs $\omega$ in Eq.\eqref{chiwk}, therefore, we can compute the QNMs with respect to $|1-E/E_c|$ while fixing $k=0$.  In the right panel of Fig.\ref{relaxationtime} we show the linear relation between the imaginary parts of the lowest modes of the QNMs and $(1-E/E_c)$. The relation is roughly  $\text{Im}(\omega)/\mu_c\approx -4.7409\times10^{-4} |1-E/E_c|$, therefore, $\tau=1/{\rm Im}(\omega)\propto |1-E/E_c|^{-1}$ (The lowest modes of $\omega$ have negative imaginary parts indicate that the system is stable against perturbations).  Hence, we get $\tau\propto\xi^2$ and $z=2$ which is the same as that in the equilibrium field theory. This also indicates that our non-equilibrium system belongs to the A-model defined in \cite{HH:1977} and satisfies $z=2-\eta$ as well. 

\section{Conclusions and Discussions}
\label{sec:cd}
We systematically studied the critical exponents of the universal scaling laws nearby the critical point of a holographic non-equilibrium phase transition, which was driven by an AC electric field sitting in the boundary of the bulk. The final states entered into a non-equilibrium steady state, rather than an equilibrium state, due to the external periodic electric field. In the final steady state, we ignored the tiny oscillations of the order parameters and took advantage of their average values in our numerics. By varying the amplitude $E$ of the applied AC electric field, we found that beyond a critical value of the amplitude $E_c$, the initial superconducting phase would be dramatically destroyed into a normal state with average vanishing condensate. We numerically calculated the six static and one dynamical critical exponents nearby this non-equilibrium phase transition critical point. It was found that these critical exponents had similar values compared to their counterparts in equilibrium dynamics, i.e., $(\alpha, \beta, \gamma, \delta, \nu, \eta)=(0,\frac12,1,3,\frac12,0)$ and $z=2$. Therefore, these exponents also satisfy the four identities \eqref{rush}, \eqref{wido}, \eqref{fish} and \eqref{jose} as shown in the Appendix \ref{sec:app}. This result was consistent with previous studies in holography that the holographic superconductors behave as a mean-field theory. The reason may be that the scaling laws nearby the critical point is a large scale behavior, which ignores the short-range or quantum properties of the system. Besides, the large $N_c$ limit of the AdS/CFT correspondence will suppress the quantum effect of the boundary field theory although it is strongly coupled.  Therefore, in the long-range limit the universal scaling laws look alike between the equilibrium and non-equilibrium dynamics.

In this paper we only considered the effects of the amplitude $E$ to the phase transition as well as the critical exponents. It will be interesting to see whether the frequency $\Omega$ will have similar effects to the critical exponents. Indeed, from our tentative computation we find that the condensate of the order parameter will behave similarly to that in mean field theory, such as $\langle O_f\rangle \propto |\epsilon_\kappa|^{1/2}$, where $\epsilon_\kappa=1-\kappa/\kappa_c$ and $\kappa\equiv\mu_c/\Omega$. Therefore, one can deduce that the static critical exponent $\beta=1/2$ if we regard frequency as a controlling parameter. The interesting thing is that from the above relation $\langle O_f\rangle \propto |\epsilon_\kappa|^{1/2}$ one finds that if $\Omega>\Omega_c$ (where $\Omega_c=\mu_c/\kappa_c$) the system will remain in the superconducting phase while on the contrary $\Omega<\Omega_c$ they system will be in the normal state with vanishing condensate. This counter-intuitive phenomenon actually can be explained by the ``Wyatte-Dayem" effect in condensed matter physics \cite{WD}, where higher frequency fields will enhance the superconductivity \cite{Eliashberg,Tinkham}. From the tentative result $\beta=1/2$ we expect that other critical exponents, such as $(\alpha,\gamma, \delta, \nu, \eta)$ and $z$ will have similar values to those in mean field theory.  We will leave this interesting topic as a future work.

\section*{Acknowledgments}
HBZ and HQZ are supported by the National Natural Science Foundation of China (Grant No. 11675140, 11705005, 11875095). 

\appendix
\section{Power-law Scaling and Critical Exponents in Equilibrium Dynamics }
\label{sec:app}

In the equilibrium dynamics, there are six static critical exponents $(\alpha, \beta, \gamma, \delta, \nu, \eta)$ and one dynamical critical exponent $z$. The static critical exponents, such as in the ferromagnetic phase transition, can be obtained from the following definitions \cite{itzykson}: 
\begin{eqnarray}
&&C\propto|\epsilon_T|^{-\alpha},\quad M\propto|\epsilon_T|^{\beta},\quad \chi\propto|\epsilon_T|^{-\gamma},\nonumber\\
 &&M\propto|h|^{1/\delta}, \quad \chi\propto e^{-r/\xi}\propto r^{2-d-\eta}\propto k^{\eta-2},\quad \xi\propto|\epsilon_T|^{-\nu}. 
\end{eqnarray} 
 In which, $\epsilon_T=(T_c-T)/T_c$ is the reduced temperature;  $C$ is the heat capacity; $M$ is the magnetization; $\chi$ is the static susceptibility; $h$ is the external magnetic field; $d$ is the spatial dimension; $k$ is the momentum of the modes and $\xi$ is the correlation length of the order parameter. In the mean-field theory, the six critical exponents satisfy the following relations:
\begin{eqnarray}
&&\alpha+2\beta+\gamma=2~~~~\rm{(Rushbrooke)}, \label{rush}\\
&&\gamma=\beta(\delta-1)~~~~\rm{(Widom)}, \label{wido}\\
&&\gamma=\nu(2-\eta)~~~~\rm{(Fisher)}, \label{fish}\\
&&2-\alpha=d\nu~~~~\rm{(Josephson)}.\label{jose} 
\end{eqnarray}
Normally,  the critical exponents in the equilibrium dynamics are $(\alpha, \beta, \gamma, \delta, \nu, \eta)=(0,\frac12,1,3,\frac12,0)$. 

In the dynamical case, we will only focus on the model A equilibrium system. From  \cite{HH:1977}, the dynamical exponents $z$ can be obtained from $\tau\propto\xi^z$ where $\tau$ is the relaxation time. For the model A $z=2-\eta$.

\end{document}